\documentclass[journal]{IEEEtran}
\usepackage{stfloats}
\usepackage{float}
\usepackage{graphicx}
\usepackage[caption=false,font=normalsize,labelfont=sf,textfont=sf]{subfig}
\usepackage{color}
\usepackage{algpseudocode}
\usepackage{amssymb}
\usepackage{amsmath}
\usepackage{algorithm}
\usepackage{bm}
\usepackage[colorlinks,citecolor=blue,linkcolor=blue,urlcolor=blue]{hyperref}

\newtheorem{remark}{Remark}

\begin{document}
\title{\LARGE Meta Fluid Antenna: Architecture Design, Performance Analysis, Experimental Examination}

\author{Baiyang Liu,  
	Jiewei Huang, 
	Tuo Wu, 
	Huan Meng,
	 Fengcheng Mei,
	  Lei Ning,  \\
	 Kai-Kit Wong, \textit{Fellow, IEEE}, 
Hang Wong, \textit{Fellow, IEEE},
 Kin-Fai Tong, \textit{Fellow, IEEE},\\ 
 and Kwai-Man Luk, \IEEEmembership{Life Fellow,~IEEE}
 
 \vspace{-5mm}
  
 \thanks{This research work of K. F. Tong was partly funded by Hong Kong Research Grants Council under the Area of Excellence Scheme under Grant AoE/E-101/23-N and the Hong Kong Metropolitan University, Staff Research Startup Fund: FRSF/2024/03. The work of T. Wu, H. Wong and K. M. Luk was funded by Hong Kong ResearchGrants Council under the Area of Excellence Scheme under Grant AoE/E-101/23-N.}
 	 \thanks{B. Liu, J. Huang,  H. Meng, F. Mei and L. Ning are with the College of Big Data and Internet, Shenzhen Technology University, Shenzhen, China (E-mail: $\rm \{liubaiyang, huangjiewei, meifengcheng, menghuan, ninglei\}$ $\rm @sztu.edu.cn$).  T. Wu,  H. Wong, and  K.-M. Luk are with the State Key Laboratory of Terahertz and Millimeter Waves, City University of Hong Kong, Hong Kong, China (E-mail: $\rm \{tuowu2, hang.wong, eekmluk\}@cityu.edu.hk$). K.-K. Wong is with the Department of Electronic and Electrical Engineering, University College London, London, UK (E-mail:$\rm  kai$-$\rm kit.wong@ucl.ac.uk$). K.-F. Tong is with the School of Science and Technology, Hong Kong Metropolitan University, Hong Kong, China (E-mail: $\rm  ktong@hkmu.edu.hk$).
 	  
 	 \textit{Baiyang Liu, Jiewei Huang, and Tuo Wu contributed equally to this work.}
 	 
 	 (\textit{Corresponding author: Kin-Fai Tong.})} 
 	}
% each address must have a unique identifier in the option field

\markboth{Prepared for submission to IEEE Transactions on Wireless Communications}{}

\maketitle

\begin{abstract}   
Fluid antenna systems (FAS) have recently emerged as a promising solution for sixth-generation (6G) ultra-dense connectivity. These systems utilize dynamic radiating and/or shaping techniques to mitigate interference and improve spectral efficiency without relying on channel state information (CSI). The reported improvements achieved by employing a single dynamically activated radiating position in fluid antenna multiple access (FAMA) are significant. To fully realize the potential of FAMA in multi-user multiplexing, we propose leveraging the unique fast-switching capabilities of a single radio-frequency (RF)-chain meta-fluid antenna structure to achieve multi-activation. This allows for a significantly larger set of independent radiating states without requiring additional signal processing. Simulations demonstrate that multi-activation FAMA enables robust multi-user multiplexing with a higher signal-to-interference ratio (SIR) under various Rayleigh-fading environments compared to other single RF-chain technologies. We further show that the SIR can be optimized within a 15~$\mu s$ timeframe under a multi-user Rayleigh-fading channel, making the proposed scheme highly suitable for fast-changing wireless environments. Verified through the theoretical Jakes' model, full three-dimensional (3D) electromagnetic (EM) simulations and experimental validation, multi-activation FAMA enables effective CSI-free, multi-user communication, offering a scalable solution for high-capacity wireless networks.
\end{abstract}


 \vspace{-3mm}
\section{Introduction}
\IEEEPARstart{T}{he} sixth-generation (6G) wireless communication systems are anticipated to meet unprecedented requirements for high data rates, ultra-low latency, and spectral efficiency exceeding 1000 bits per second per Hertz (bps/Hz) to support emerging applications such as immersive extended reality, holographic communications, and autonomous systems~\cite{you2023toward,khalid2021advanced,wang2023full,elmeadawy2020enabling}. Conventional multiple-input multiple-output (MIMO) architectures, despite their proven effectiveness in current networks, exhibit inherent limitations in scalability and adaptability for 6G scenarios due to their reliance on fixed physical structures with limited reconfiguration capabilities.

Fluid antenna systems (FAS) have recently emerged as a paradigm-shifting technology that addresses these fundamental limitations through physically reconfigurable antenna structures~\cite{new2024tutorial,wong2020fluid,wong2021fluid}. By enabling dynamic port selection and position optimization within a constrained physical space, FAS fundamentally decouples the radiation characteristics from fixed hardware constraints, thereby facilitating adaptive spatial multiplexing and diversity exploitation. Recent investigations have demonstrated FAS's potential advantages in various aspects of wireless communications, including capacity enhancement, reliability improvement, and resource utilization optimization~\cite{lai2024fasris,yao2025fasris,xu2023channel,ghadi2023copula,new2023information,waqar2023deep,wang2024fluid}.

To realize the concept of FAS and achieve their envisioned functionalities, various architectures and prototypes have been developed in recent years~\cite{liu2025programmable}. \textbf{Liquid metal-based FAS} represent the earliest approach, utilizing conductive liquid metals (such as Galinstan alloy) that can be moved to different ports through microfluidic channels, enabling single RF-chain multi-antenna effects \cite{wong2020fluid,wong2023fluid,wu2024FAS}. However, the mechanical control of liquid metal movement is   slow,  which is inadequate for   high-mobility 6G scenarios. \textbf{Mechanically movable FAS} offers another approach by physically repositioning antenna elements, but they suffer from even slower reconfiguration speeds and mechanical wear issues that limit their practical deployment \cite{new2024tutorial,wong2020fluid1}. \textbf{Reconfigurable pixel-based FAS} emerged as a promising solution to overcome the speed limitations of mechanical approaches. These systems employ electronically controlled switching elements (such as PIN diodes) to modify the connectivity between antenna pixels, effectively changing the antenna's shape and radiation characteristics \cite{zhang2024pixel,zhang2025novel}. Each reconfigurable state corresponds to a different antenna geometry, enabling microsecond-scale switching speeds suitable for real-time adaptation. However, pixel-based FAS face two critical limitations: \textbf{(i) Limited reconfigurable states} - the number of distinct radiation patterns is constrained by the finite switching combinations, restricting the spatial diversity that can be achieved; \textbf{(ii) Poor electric field independence} - the radiation patterns of different reconfigurable states exhibit high correlation, leading to decreased signal-to-interference ratio (SINR) performance and reduced diversity gain. Additionally, most pixel-based designs operate at relatively low frequency bands, limiting their applicability to mmWave 6G systems.

To transcend these limitations of existing FAS prototypes, we introduce the \textbf{meta-fluid antenna}, a revolutionary antenna architecture that achieves genuine fluid-like electromagnetic behavior through advanced reconfigurable slot antenna array engineering. The fundamental innovation of M-FAS lies in its ability to achieve \textbf{microsecond-scale pseudo-fluid dynamics} through electronically controlled meta-atom elements. Each meta-atom can be dynamically reconfigured to alter its electromagnetic properties, enabling the antenna to reshape its radiation pattern and effective aperture in real-time, creating continuous, microsecond-scale electromagnetic field redistribution that truly mimics the flow characteristics of physical fluids. \textbf{By enabling dynamic control over electromagnetic field distribution at the meta-atom level, M-FAS effectively expands the spatial degrees of freedom (DoF), creating a virtual aperture that surpasses the physical limitations of traditional antenna arrays}. This spatial DoF expansion directly translates to enhanced communication performance metrics, including improved channel capacity, increased signal-to-interference ratio, and more effective interference mitigation across diverse propagation environments while maintaining the flexibility and adaptability essential for 6G networks.

The meta-fluid antenna architecture enables a revolutionary multiple access paradigm,  \textbf{meta-fluid antenna multiple access (M-FAMA)}~\cite{wong2023slow,zhang2025ura},  that leverages the unique fluid-like electromagnetic properties for interference mitigation and capacity enhancement~\cite{wong2023compact,xu2024capacity}. Unlike conventional  multiple access techniques that requires multiple RF chains for multi-position activation, M-FAMA achieves superior performance through a single RF chain by exploiting the continuous electromagnetic field manipulation capabilities of the meta-fluid structure. This approach fundamentally differs from massive multi-user multiple-input multiple-output (MU-MIMO) \cite{urquiza2022spectral,wong2002performance,vishwanath2003duality,spencer2004introduction,wong2000optimizing,spencer2004zero,choi2004transmit,gesbert2007shifting}, non-orthogonal multiple access (NOMA)~\cite{ding2017survey,new2024enhancing,yao2025fairness,zheng2024shortpacket}, and rate-splitting multiple access (RSMA)~\cite{mao2022rate,ghadi2025rsma} schemes by eliminating the need for channel state information (CSI) feedback and complex precoding/decoding operations while providing continuous spatial adaptation capabilities, positioning M-FAS as a superior solution for next-generation wireless communications requiring both high performance and rapid adaptability.

However, achieving these ambitious goals through M-FAS implementation presents significant technical challenges. To overcome \textbf{(i) limited reconfigurable states}, M-FAS must provide virtually unlimited radiation pattern combinations through its 120-element meta-atom array, requiring sophisticated control of 480 PIN diodes (four per meta-atom) with precise timing coordination via FPGA-based control systems operating at 20 MHz clock rates to achieve sub-15 $\mu$s reconfiguration times. To address \textbf{(ii) poor electric field independence}, M-FAS must ensure that different activation patterns produce truly uncorrelated radiation patterns, demanding careful electromagnetic design of slot element geometries, phase relationships, and substrate-integrated waveguide (SIW) feeding structures to maintain high power contrast (larger than 10 dB) between radiating and non-radiating states across all 120 positions. To enable \textbf{(iii) mmWave compatibility}, M-FAS   operates effectively in 6G FR2 frequency band, requiring precise fabrication tolerances, thermal stability management, and robust PIN diode performance at millimeter-wave frequencies where parasitic effects become critical.

The practical realization of these capabilities introduces additional implementation challenges. The meta-atom-based reconfiguration demands real-time coordination of hundreds of switching elements while maintaining electromagnetic coherence across the entire aperture. The continuous electromagnetic field manipulation requires sophisticated algorithms capable of optimizing activation patterns within microsecond timeframes. Furthermore, the dynamic nature of M-FAS necessitates efficient performance analysis frameworks that can characterize the continuously varying electromagnetic behavior across diverse channel conditions.

To address these multifaceted challenges, this paper presents a comprehensive M-FAS design and analysis framework that systematically tackles the theoretical, algorithmic, and experimental complexities associated with meta-fluid antenna implementation. We formulate the M-FAS optimization problem to achieve optimal meta-atom activation patterns through real-time electromagnetic field manipulation. We derive analytical expressions for system performance under spatial correlation effects. We develop fast heuristic algorithms capable of optimizing activation patterns within microsecond timeframes while maintaining electromagnetic coherence across the 120-element array. We validate our theoretical framework through comprehensive electromagnetic simulations and experimental measurements at 26.5 GHz, demonstrating substantial performance improvements in multi-user scenarios with notable enhancements in spectral efficiency and interference management. This work's primary contributions are summarized as follows:

\begin{itemize}
\item \textbf{\textit{Revolutionary Meta-Fluid Antenna Architecture}}---We introduce the first M-FAS that achieves true electromagnetic fluid behavior through advanced reconfigurable slot array engineering. Unlike conventional FAS approaches that rely on mechanical switching or discrete position selection, our architecture employs 120 electronically controlled meta-atom elements with PIN diode switching to create continuous, microsecond-scale electromagnetic field redistribution. We specifically develop a novel substrate-integrated waveguide (SIW) feeding structure that supports single RF chain operation while maintaining 13 dB power contrast between radiating and non-radiating states across all positions.

\item \textbf{\textit{Single-Chain Multi-Activation Protocol}}---We develop a revolutionary M-FAMA protocol that leverages meta-fluid properties to achieve multi-position activation effects using only a single RF chain. Our approach fundamentally differs from conventional multi-chain FAS by exploiting continuous electromagnetic field manipulation capabilities, dramatically reducing hardware complexity while eliminating the need for channel state information (CSI) feedback and complex precoding/decoding operations.

\item \textbf{\textit{Comprehensive Theoretical Framework}}---We establish a rigorous analytical framework for M-FAS performance analysis. Our analysis provides tractable mathematical tools that capture the unique characteristics of meta-fluid electromagnetic behavior, enabling efficient system design and optimization while maintaining high accuracy for compact deployments with limited ports ($N<20$).

\item \textbf{\textit{Experimental Validation and Performance Demonstration}}---We present the first experimental demonstration of a meta-fluid antenna system operating at 26.5 GHz, achieving sub-15 $\mu$s reconfiguration times through FPGA-based control with 20 MHz system clock. Our comprehensive validation includes full-wave electromagnetic simulations, theoretical analysis using Jakes' model, and real-world measurements, confirming the practical feasibility of M-FAS and validating theoretical predictions across diverse operating conditions.

\item \textbf{\textit{Practical Implementation Insights}}---Our analysis reveals that M-FAS can achieve substantial performance improvements with enhanced spatial degrees of freedom and superior channel utilization efficiency. We provide practical guidelines for meta-atom design, PIN diode control strategies, and system optimization that can inform real-world 6G network deployments requiring both high performance and rapid adaptability.
\end{itemize}

 %The remainder of this paper is organized as follows. Section~\ref{sec:architecture} introduces the meta-fluid antenna architecture and prototype design, detailing the reconfigurable slot array concept and implementation. Section~\ref{sec:system_model} presents the system model and channel characterization, including spatial correlation modeling and block-correlation approximation. In Section~\ref{sec:analytical_performance}, we develop the analytical performance analysis framework with PDF and CDF derivations for outage probability evaluation. Section~\ref{sec:multi_activation_fama} details the multi-activation FAMA protocol with enhanced correlation modeling. Afterwards, Section~\ref{sec:simulation} provides comprehensive simulation results and analytical framework validation. Finally, the experimental validation is presented, demonstrating the practical feasibility of the proposed M-FAS system and confirming theoretical predictions through real-world measurements.
 \vspace{-3mm}
\section{Meta-Fluid Antenna Architecture and Prototype Design}\label{sec:architecture}

\subsection{Reconfigurable Slot Array-Based Fluid Antenna Concept}
The meta-fluid antenna represents a fundamental departure from conventional reconfigurable antenna designs by achieving true electromagnetic fluid behavior through advanced reconfigurable slot array engineering. Unlike traditional fluid antenna systems that rely on mechanical switching or discrete position selection, our M-FAS employs dynamically reconfigurable slot elements with PIN diode control to create continuous, microsecond-scale electromagnetic field redistribution across the antenna aperture.

The core innovation lies in the multi-activation of electronically controlled meta-atoms units, each comprising multiple slot elements and PIN diodes capable of dynamic switching to alter the radiating characteristics in real-time. This enables the antenna to shift its radiation pattern and utilize  effective aperture continuously, creating an electromagnetic ``fluid" that can flow and redistribute across the antenna surface within 15 $\mu$s timeframes. Such rapid reconfiguration capability is orders of magnitude faster than mechanical approaches, making it particularly suitable for high-mobility 6G scenarios.
\begin{figure}
	\centering
	\includegraphics[width=9cm]{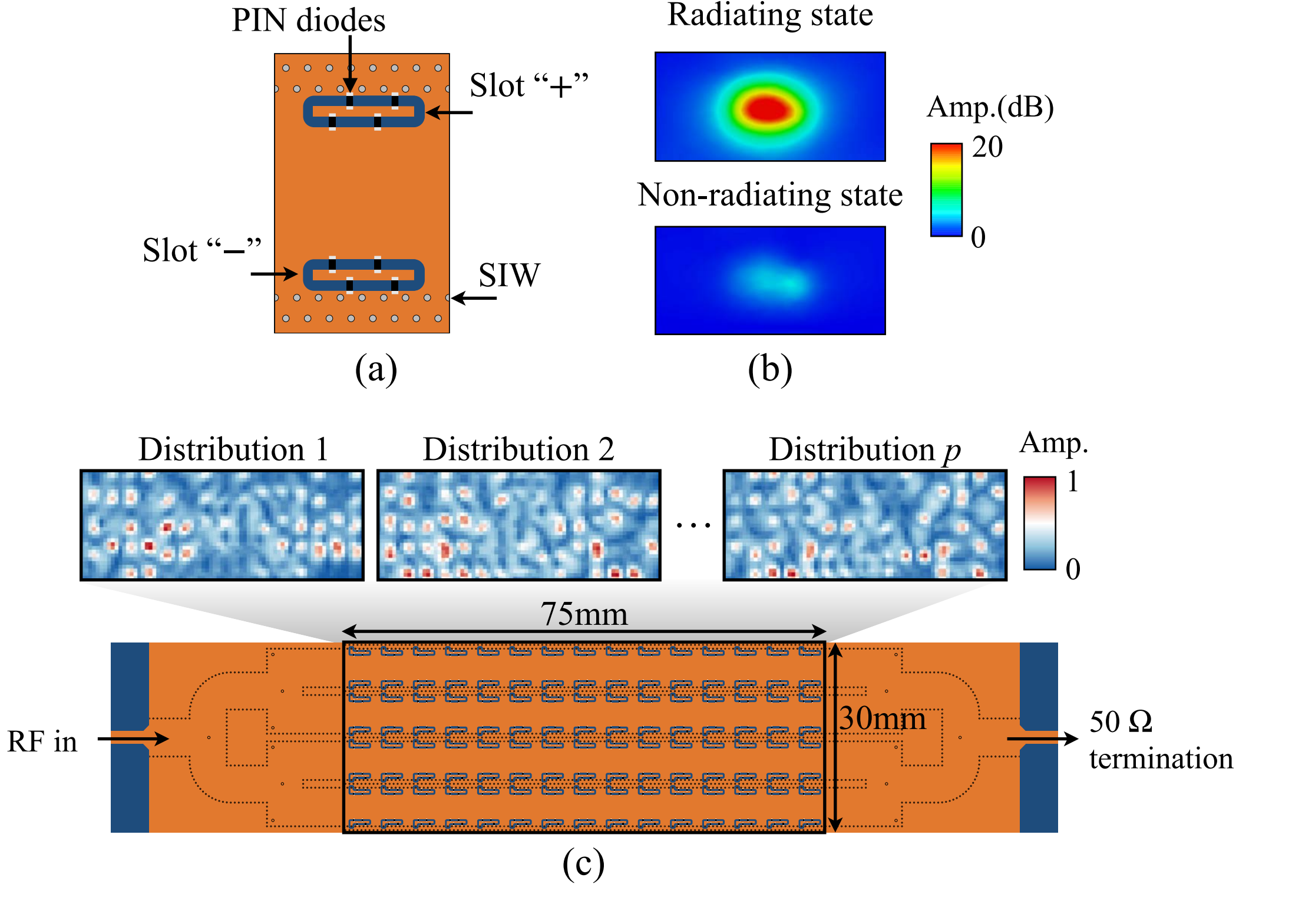}
	\caption{(a) The complete structure of the meta-atom  pair comprise  two slot elements, designated as slot ``\(+\)" and slot
		``\(-\)", which are controlled by the operating states of the PIN diodes. The key distinction between the two slots lies in
		their phase difference. (b) The difference in the electric field (E-field) between the radiating and non-radiating states
		demonstrates a power contrast of 13 dB. By dynamically activating different radiating elements, it is possible to achieve a shift in the radiating position, mimicking the behavior of a fluid state. This approach allows for flexible and adaptable control over the radiation pattern, similar to how a fluid can change its form and flow. The size of the
		proposed meta-atom is 0.4 \(\lambda_{c}\) in width and 0.7\(\lambda_{c}\) in length. (c) The complete meta-fluid antenna consisting of 120 reconfigurable elements, along with the E-field distributions above the antenna under different activation patterns, illustrating the dynamic fluid-like behavior of the E-fields.
	} \vspace{-3mm}
	\label{fig:meta-fluid antenna}
\end{figure}
\subsection{Prototype Design and Implementation}
In this study, we introduce a practical and efficient meta-fluid antenna structure that incorporates PIN diodes to modulate the radiating state of each slot-based meta-atoms. The prototype design enables each individual meta-atom to be dynamically switched between radiating and non-radiating states, allowing for versatile selection of radiating elements and precise control over the electromagnetic field distribution.

Our prototype consists of a two-dimensional array with $N = 120$ reconfigurable meta-atom elements, arranged in $I = 8$ rows and $J = 15$ columns. It is worth mentioning that $N$, $I$ and $J$ can be changed for different application scenarios. Each meta-atom pair  comprises two slot elements with different phase characteristics, and the radiating state of each meta-atom can be dynamically toggled through electronic control. The system is powered by a substrate-integrated waveguide (SIW) that effectively supports 120 programmable positions through a single RF chain, dramatically reducing hardware complexity compared to conventional multi-chain approaches.

As illustrated in Fig.~\ref{fig:meta-fluid antenna}(a), each meta-atom element is equipped with four PIN diodes, which enable selective activation of the slot elements to radiate energy from the waveguide into free space, while the remaining elements remain non-radiating. The E-field amplitude difference between the radiating and non-radiating states exceeds 13 dB, facilitating the dynamic activation of specific positions for signal reception. The complete 3D electromagnetic model, including the E-field distributions above the antenna under different activation patterns, is shown in Fig.~\ref{fig:meta-fluid antenna}(c).

\subsection{Electromagnetic Characteristics and Performance Metrics}
%The electromagnetic characteristics of our meta-fluid antenna prototype demonstrate the effectiveness of the reconfigurable slot array-based approach. The reflection coefficients for 1000 different random activation distributions are presented in Fig.~\ref{fig:s11}, showing consistent performance across diverse configurations. Similarly, the radiation patterns for these distributions, shown in Fig.~\ref{fig:radiationpattern}, demonstrate the antenna's ability to achieve diverse spatial radiation characteristics through electronic reconfiguration of the slot elements.

A key feature of our design is its ability to generate controlled electromagnetic diversity through strategic activation of meta-atom elements. By strategically activating radiating slot elements in the meta-fluid antenna, disordered electromagnetic fields can be generated \cite{liu2024multifunctional}, effectively mimicking Rayleigh-fading environments in wireless communication. This capability enables the system to create distinct channel realizations for performance evaluation and optimization, with each channel represented by a unique configuration of activated slot elements.

The prototype operates at 26.5 GHz, a key frequency band for future 6G networks, and achieves sub-15 $\mu$s reconfiguration times through FPGA-based control with a 20 MHz system clock. This rapid reconfiguration capability, combined with the single RF chain architecture, positions the meta-fluid antenna as a practical solution for next-generation wireless systems requiring both high performance and low complexity.

\begin{remark}[Key Technical Advantages of M-FAS over Existing FAS Prototypes]
Unlike the aforementioned FAS prototypes that depend on mechanical reconfiguration mechanisms or suffer from limited reconfigurable states and high pattern correlation, M-FAS addresses the fundamental limitations through three key technical breakthroughs: \textbf{(i) Ultra-fast reconfiguration capability} - achieving sub-15 microsecond switching times through electronic control rather than mechanical movement, orders of magnitude faster than mechanical reconfiguration approaches; \textbf{(ii) Enhanced pattern diversity} - the radiation patterns of M-FAS exhibit significantly lower correlation compared to pixel-based antennas, thereby substantially increasing the achievable SINR and diversity gain; \textbf{(iii) mmWave compatibility} - operating effectively at 26.5 GHz and higher frequencies critical for 6G deployment. Such rapid reconfiguration enables the system to track and adapt to fast-changing wireless environments, making it particularly suitable for high-mobility 6G scenarios where conventional antennas fail to maintain optimal performance.
\end{remark}
 \vspace{-1mm}
\section{System Model and Channel Characterization}\label{sec:system_model}

\subsection{System Configuration}
Building upon the meta-fluid antenna architecture described in Section~\ref{sec:architecture}, we now establish the mathematical framework for system-level analysis. Consider a downlink system where the base station (BS) is equipped with \( N_t \) conventional fixed antennas, and there are \( U \) users, each employing the meta-fluid antenna system (M-FAS) introduced in the previous section.

Each M-FAS consists of the two-dimensional reconfigurable slot array with $N_1$ meta-atom elements uniformly distributed along the first dimension of length $W_1 \lambda$ and $N_2$ elements along the second dimension of length $W_2 \lambda$, where \(\lambda\) represents the wavelength. As detailed in Section~\ref{sec:architecture}, this configuration provides a total of $N = N_1 \times N_2 = 120$ electronically switchable meta-atom elements within an overall aperture of dimensions $W_t = W_1 \lambda \times W_2 \lambda$, with each element capable of dynamic activation through PIN diode control.

\subsection{Signal Model}
In this system, each transmitting antenna at the BS is dedicated to a specific user, transmitting signals intended for that user. At each user, the M-FAS dynamically selects and activates a subset of its $N$ available meta-atom elements for signal reception\footnote{All activated elements within an M-FAS share a single RF chain through the substrate-integrated waveguide structure, significantly reducing hardware complexity compared to conventional multi-antenna systems. This is achieved through our specific meta-fluid antenna design detailed in Section~\ref{sec:architecture}.}. The received signal at the $u$-th user is expressed as:
\begin{equation}\label{eq:signal_model}
    \mathbf{r}_{u} =  \mathbf{H}_{u} s_u + \sum_{u' \neq u}^{U} \mathbf{H}_{u'} s_{u'} + \boldsymbol{\eta}_{u},
\end{equation} 
where $\mathbf{H}_{u} \in \mathbb{C}^{N \times 1} = [h_{u}^{1}, h_{u}^{2}, \ldots, h_{u}^{N}]^T$ represents the channel vector from the transmitting antenna assigned to user $u$ to all $N$ meta-atom elements of user $u$'s M-FAS, $s_u \sim \mathcal{CN}(0, \sigma_s^2)$ denotes the data symbol transmitted to user $u$, and $\boldsymbol{\eta}_{u} \in \mathbb{C}^{N \times 1} = [\eta_{u}^{1},\eta_{u}^{2},\ldots,\eta_{u}^{N}]^T$ with $\eta_{u}^{i} \sim \mathcal{CN}(0, \sigma_n^2)$ representing the additive white Gaussian noise (AWGN) at each meta-atom element.

\subsection{Spatial Correlation Modeling}
For analytical simplicity, the 2D meta-atom element indices are mapped to one-dimensional (1D) indices, sequentially numbered from left to right and top to bottom:
\begin{equation}\label{eq:mapping}
m(n_1, n_2) = (n_2 - 1) N_1 + n_1,
\end{equation}
where $n_1 \in [1, N_1]$ and $n_2 \in [1, N_2]$ represent the row and column indices of the reconfigurable slot array described in Section~\ref{sec:architecture}. 

Considering a 3D environment under rich scattering, the spatial correlation matrix between meta-atom elements follows Jakes' model \cite{stuber2002principles}:
\begin{equation}\label{eq:jakes_corr}
	\begin{aligned}
	\boldsymbol{\Sigma}_{m(n_1, n_2), m(\tilde{n}_1, \tilde{n}_2)} &= \text{Cov}(h_{u}^{m(n_1, n_2)}, h_{u}^{m(\tilde{n}_1, \tilde{n}_2)}) \\
	&= \sigma^2 J_0\left(2\pi \left|\Delta d_{(n_1, n_2), (\tilde{n}_1, \tilde{n}_2)}\right|/\lambda\right),
	\end{aligned}
\end{equation}
where $\Delta d_{(n_1, n_2), (\tilde{n}_1, \tilde{n}_2)} = \sqrt{\left(\frac{|n_1 - \tilde{n}_1|}{N_1 - 1} W_1 \lambda \right)^2 + \left(\frac{|n_2 - \tilde{n}_2|}{N_2 - 1} W_2 \lambda \right)^2}$ represents the physical distance between meta-atom elements in the array.

\subsection{Block-Correlation Approximation}
The Toeplitz-structured correlation matrix $\boldsymbol{\Sigma} \in \mathbb{C}^{N \times N}$ from \eqref{eq:jakes_corr} makes direct analytical treatment intractable. To enable tractable outage probability analysis, we adopt a block-correlation approach that approximates $\boldsymbol{\Sigma}$ with a block-diagonal structure:
\begin{equation}\label{eq:blockdiag}
\hat{\boldsymbol{\Sigma}} = \mathrm{blkdiag}(\mathbf{A}_1, \mathbf{A}_2, \ldots, \mathbf{A}_D),
\end{equation}
where each block $\mathbf{A}_d \in \mathbb{C}^{L_d \times L_d}$ has the form:
\begin{equation}\label{eq:block}
\mathbf{A}_d = \begin{bmatrix}
1 & \rho_d & \cdots & \rho_d \\
\rho_d & 1 & \cdots & \rho_d \\
\vdots & \vdots & \ddots & \vdots \\
\rho_d & \rho_d & \cdots & 1
\end{bmatrix},
\end{equation}
where $D$ is the number of blocks, $\rho_d$ is the correlation coefficient for the $d$-th block, and $\sum_{d=1}^D L_d = N$.

\subsubsection{Constant Block-Correlation Model}
First we can use a constant correlation coefficient across all blocks \cite{BCM}, i.e., $\rho_d = \rho$ for all $d \in \{1, 2, \ldots, D\}$, where $\rho \in [0.95, 0.97]$ is typically chosen empirically. The number of blocks $D$ is determined by the number of dominant eigenvalues of $\boldsymbol{\Sigma}$, and the block sizes $\{L_d\}_{d=1}^D$ are optimized to minimize:
\begin{equation}\label{eq:constant_optimization}
\min_{L_1, \ldots, L_D} \sum_{d=1}^{D} |\lambda_d - \hat{\lambda}_d|^2,
\end{equation}
where $\hat{\lambda}_d = (L_d-1)\rho + 1$ represents the maximum eigenvalue of the $d$-th block.

While this constant correlation model achieves satisfactory approximation when $N$ is large (as $\boldsymbol{\Sigma}$ typically has only a few dominant eigenvalues), it may introduce significant errors when $N$ is small. In such cases, the eigenvalues of $\boldsymbol{\Sigma}$ exhibit smaller variations, making it challenging to achieve accurate approximation with a fixed $\rho$.

\subsubsection{Variable Block-Correlation Model}
While the constant block-correlation model provides a tractable approximation for large $N$ scenarios, it exhibits significant limitations when applied to compact M-FAS deployments where $N < 20$. {The original work in \cite{BCM} recognized this limitation and proposed the concept of variable block-correlation modeling, where correlation coefficients $\rho_d$ can vary across different blocks to achieve better approximation accuracy.} However, {due to the high computational complexity of the resulting optimization problem, \cite{BCM} did not provide a specific algorithmic solution to determine the optimal variable correlation parameters.}

{To bridge this gap and enable practical implementation of variable block-correlation modeling for M-FAS systems,} we develop a comprehensive algorithmic framework that systematically determines the optimal block parameters. The variable block-correlation approach optimizes each correlation coefficient $\rho_d$ individually through the following optimization objective:
\begin{equation}\label{eq:eigmin}
\min_{\{\rho_d\}_{d=1}^D, \{L_d\}_{d=1}^D} \left\|\mathrm{eig}(\boldsymbol{\Sigma}) - \mathrm{eig}(\hat{\boldsymbol{\Sigma}})\right\|_2^2,
\end{equation}
where $\mathrm{eig}(\cdot)$ denotes the eigenvalue vector in decreasing order, and $\hat{\boldsymbol{\Sigma}}$ represents the block-diagonal approximation from \eqref{eq:blockdiag}.

{  The optimization problem in \eqref{eq:eigmin} can be decomposed into manageable subproblems through eigenvalue-based block assignment. For a given block assignment, the assignment error when eigenvalue $\lambda_{\text{cur}}$ is added to block $d$ is quantified by}
{\begin{multline}\label{eq:block_error}
\mathrm{dist}_d = (1 + \rho_d(L_d-1) - \lambda_d)^2\\
 + \sum_{k \in K_d \cup \{\lambda_{\text{cur}}\}} (\lambda_k - 1 + \rho_d)^2,
\end{multline}}
{where $K_d$ contains the eigenvalues currently assigned to block $d$, and $L_d = |K_d| + 1$ is the resulting block size.}

The optimal correlation coefficient for the $d$-th block is obtained by minimizing the block error, yielding the closed-form expression:
\begin{equation}\label{eq:optimal_rho}
\tilde{\rho}_d = \min\left(\max\left(\frac{(L_d-1)\lambda_d - \sum_{k=1}^{L_d-1}\lambda_{d,k}}{2(L_d-1)}, 0\right), 1\right),
\end{equation}
where $\lambda_d$ and $\{\lambda_{d,k}\}$ are the eigenvalues allocated to block $d$.

 To realize the variable block-correlation concept proposed in \cite{BCM}, we develop a computationally efficient eigenvalue assignment procedure. The VBCM optimization follows a systematic approach where the eigenvalues of the original correlation matrix $\boldsymbol{\Sigma}$ are sorted in descending order and split into two groups: the dominant eigenvalues $\Lambda_1=[\lambda_1,\dots,\lambda_D]$ and the remaining eigenvalues $\Lambda_2=[\lambda_{D+1},\dots,\lambda_N]$. The dominant eigenvalues are reserved as the principal eigenvalues for each of the $D$ blocks, while the remaining eigenvalues are systematically assigned to blocks using a greedy strategy that maintains linear computational complexity. 

{ For each remaining eigenvalue $\lambda_{\text{cur}}$ in $\Lambda_2$, the optimization process evaluates its assignment to each of the $D$ blocks by computing the resulting block size $L_d = |K_d| + 1$, where $K_d$ represents the set of eigenvalues currently assigned to block $d$. The optimal correlation coefficient $\tilde{\rho}_d$ is calculated using \eqref{eq:optimal_rho}, and the assignment error is determined through \eqref{eq:block_error}. The eigenvalue is then assigned to the block that minimizes this assignment error, ensuring optimal approximation while maintaining computational efficiency.}

%{\textbf{Key Contribution and Advantages:} Our algorithmic framework successfully addresses the implementation gap left by \cite{BCM} by providing a concrete solution to the variable block-correlation optimization problem. The proposed approach achieves: (i) \textit{Enhanced accuracy} - particularly effective for compact M-FAS deployments with $N < 20$, capturing subtle correlation variations that significantly impact system performance; (ii) \textit{Computational efficiency} - complexity reduced from exponential $D^{N-D}$ (exhaustive search) to linear $(N-D) \times D$ through the heuristic assignment strategy; (iii) \textit{Practical implementability} - unlike the conceptual framework in \cite{BCM}, our method provides a complete algorithmic solution that can be readily deployed in M-FAS systems.}

{Having established this practical variable block-correlation framework that realizes the concepts proposed in \cite{BCM}, we now proceed to develop the effective channel model that leverages this enhanced correlation modeling for accurate M-FAS performance analysis.}

\subsection{Effective Single-Chain Channel Model}
The M-FAS employs an element activation vector $\boldsymbol{\omega} = [\omega_1, \omega_2, \ldots, \omega_N]^T$ where $\omega_i \in \{0,1\}$ indicates whether the $i$-th meta-atom element is activated through PIN diode control as described in Section~\ref{sec:architecture}. The effective channel after element selection becomes:
\begin{equation}\label{eq:effective_channel}
z_u = \boldsymbol{\omega}^T \mathbf{H}_u = \sum_{i=1}^N \omega_i h_u^i,
\end{equation}
which represents the combined signal from all activated meta-atom elements through the single RF chain via the substrate-integrated waveguide structure. Under the block-correlation approximation \eqref{eq:blockdiag}, $z_u \sim \mathcal{CN}(0, \Omega_u)$ where:
\begin{equation}\label{eq:omega_u}
\Omega_u = \boldsymbol{\omega}^T \hat{\boldsymbol{\Sigma}} \boldsymbol{\omega}.
\end{equation}

\subsection{Performance Objective and Analysis Framework}
The primary objective of this work is to derive accurate outage probability expressions for the proposed single RF chain multi-activation M-FAS system. Our analytical framework aims to characterize the outage probability based on the SIR distribution derived from the effective channel model in \eqref{eq:effective_channel}, while analyzing how the spatial correlation matrix $\boldsymbol{\Sigma}$ (and its block-diagonal approximation $\hat{\boldsymbol{\Sigma}}$ in \eqref{eq:blockdiag}) affects outage performance. The framework provides tractable analytical expressions that enable efficient system design and optimization, directly connecting system-level performance to the enhanced correlation modeling for accurate performance prediction, particularly when $N < 20$.

To evaluate system performance, we introduce key performance metrics that leverage the enhanced correlation modeling. The outage probability, based on the effective SIR from the single RF chain, is defined as:
\begin{equation}\label{eq:outage_metric}
p_{\text{out}}(\gamma) = \Pr(\text{SIR} < \gamma),
\end{equation}
where the SIR distribution is analytically characterized using the block-correlation approximation. Additionally, the multiplexing gain, representing the spectral efficiency metric, is given by:
\begin{equation}\label{eq:multiplexing_gain}
m(\gamma) = U(1 - p_{\text{out}}(\gamma)),
\end{equation}
which quantifies the effective number of concurrent users supported at SIR threshold $\gamma$.

\begin{remark}[Distinctive Features of M-FAS]
The proposed M-FAS exhibits several distinctive characteristics that differentiate it from conventional multi-antenna systems: \textbf{(i) Single RF Chain Architecture:} Unlike traditional MIMO systems requiring multiple RF chains, M-FAS achieves multi-position activation through a single RF chain, significantly reducing hardware complexity and power consumption while maintaining spatial diversity benefits. \textbf{(ii) Dynamic Position Reconfiguration:} The system can dynamically switch between different activation patterns within microseconds, enabling rapid adaptation to channel variations without requiring channel state information feedback. \textbf{(iii) Correlation-Aware Design:} The close proximity of antenna positions introduces significant spatial correlation, which is explicitly modeled and leveraged for performance optimization rather than being treated as a limitation. \textbf{(iv) Scalable Performance:} The system performance scales gracefully with the number of available positions $N$, particularly effective for compact deployments where $N < 20$, making it suitable for size-constrained applications in 6G networks.
\end{remark}
 \vspace{-3mm}
\section{Analytical Performance Analysis}
\label{sec:analytical_performance}

Having established the system model in Section~\ref{sec:system_model}, we now develop an independent theoretical framework for performance analysis of M-FAS systems.

\subsection{Effective Channel Statistics under Block-Correlation}
To achieve the analytical objectives outlined above, we begin by leveraging the variable block-correlation approximation from \eqref{eq:blockdiag} to analyze the statistical properties of the effective single-chain channel. {The optimized block-diagonal matrix $\hat{\boldsymbol{\Sigma}}$ obtained from the eigenvalue assignment procedure provides the foundation for accurate channel statistics analysis.} The effective desired signal for user $u$ after position activation is given by $z_u = \boldsymbol{\omega}^T \mathbf{H}_u$ from \eqref{eq:effective_channel}, where $\mathbf{H}_u \sim \mathcal{CN}(\mathbf{0}, \hat{\boldsymbol{\Sigma}})$. Since $z_u$ is a linear combination of jointly Gaussian variables, the effective signal distribution can be written as
\begin{equation}\label{eq:effective_signal_dist}
z_u \sim \mathcal{CN}(0, \Omega_u), \quad \text{where} \quad \Omega_u = \boldsymbol{\omega}^T \hat{\boldsymbol{\Sigma}} \boldsymbol{\omega}.
\end{equation}

{ The block-diagonal structure $\hat{\boldsymbol{\Sigma}} = \mathrm{blkdiag}(\mathbf{A}_1, \mathbf{A}_2, \ldots, \mathbf{A}_D)$ obtained from the eigenvalue assignment procedure enables efficient computation of $\Omega_u$.} For activation pattern $\boldsymbol{\omega}$, the variance calculation becomes
\begin{equation}\label{eq:block_variance_calc}
{\Omega_u = \sum_{d=1}^{D} \boldsymbol{\omega}_d^T \mathbf{A}_d \boldsymbol{\omega}_d},
\end{equation}
where $\boldsymbol{\omega}_d$ represents the subvector of $\boldsymbol{\omega}$ corresponding to block $d$. This decomposition significantly reduces computational complexity while maintaining high accuracy through the optimized correlation coefficients $\{\rho_d^*\}_{d=1}^D$ determined by the algorithm.  

Similarly, for each interfering user $u' \neq u$, the effective interference signal is $z_{u'} = \boldsymbol{\omega}^T \mathbf{H}_{u'} \sim \mathcal{CN}(0, \Omega_u)$ under the assumption of i.i.d. user channels and identical activation patterns. The signal and interference powers are characterized by
\begin{equation}\label{eq:power_distributions}
\begin{aligned}
X &\triangleq |z_u|^2 \sim \mathrm{Exp}(\Omega_u), \\
Y &\triangleq \sum_{u' \neq u} |z_{u'}|^2 \sim \mathrm{Gamma}(U-1, \Omega_u),
\end{aligned}
\end{equation}
where $\mathrm{Exp}(\theta)$ denotes the exponential distribution with scale parameter $\theta$, and $\mathrm{Gamma}(k, \theta)$ denotes the gamma distribution with shape parameter $k$ and scale parameter $\theta$.

The effective noise after position combining is $n_{\text{eff}} = \boldsymbol{\omega}^T \boldsymbol{\eta}_u \sim \mathcal{CN}(0, \sigma_n^2 \|\boldsymbol{\omega}\|_2^2)$. We define the normalized noise-to-signal ratio and the SIR random variable as
\begin{equation}\label{eq:sir_definition}
c \triangleq \frac{\sigma_n^2}{\sigma_s^2} \|\boldsymbol{\omega}\|_2^2, \qquad S \triangleq \frac{X}{Y + c},
\end{equation}
such that the outage probability becomes $P_{\text{out},u} = \Pr(S < \gamma_{\text{th}})$.

\subsection{Closed-Form PDF and CDF Derivation}

With the statistical characterization of signal and interference powers established in \eqref{eq:power_distributions}, we now derive closed-form expressions for the PDF and CDF of the SIR random variable $S$ defined in \eqref{eq:sir_definition}. The derivation considers two distinct operating regimes that commonly arise in practical M-FAS deployments.

\subsubsection{Interference-Limited Regime ($c \approx 0$)}
When noise is negligible compared to interference, $S = X/Y$ follows a beta-prime distribution. The CDF can be expressed in closed form as
\begin{equation}\label{eq:cdf_interference_limited}
F_S(t) = 1 - (1 + bt)^{-k}, \quad b \triangleq 1, \quad k = U-1, \quad t \geq 0,
\end{equation}
with the corresponding PDF given by
\begin{equation}\label{eq:pdf_interference_limited}
f_S(t) = kb(1 + bt)^{-(k+1)}, \quad t \geq 0.
\end{equation}

\subsubsection{General Case with Noise ($c > 0$)}
For the general case including noise, we use the independence of signal and interference powers along with the Laplace transform approach. The CDF is expressed as
\begin{equation}\label{eq:cdf_general}
F_S(t) = 1 - e^{-at}(1 + bt)^{-k}, \quad a \triangleq \frac{c}{\Omega_u}, \quad t \geq 0,
\end{equation}
with the corresponding PDF written as
\begin{equation}\label{eq:pdf_general}
f_S(t) = e^{-at}(1 + bt)^{-(k+1)}[a(1 + bt) + kb], \quad t \geq 0.
\end{equation}

\subsection{Final Outage Probability Expression}
The outage probability at threshold $\gamma_{\text{th}}$ is obtained by evaluating the CDF, which can be written as
\begin{equation}\label{eq:final_outage}
p_{\text{out}}(\gamma_{\text{th}}) = F_S(\gamma_{\text{th}}),
\end{equation}
where \eqref{eq:cdf_interference_limited} is used for interference-limited scenarios and \eqref{eq:cdf_general} for the general case.

\subsection{Analysis of PDF and CDF Characteristics}
Having derived the analytical expressions for the SIR distribution, we now provide insights into the mathematical structure and practical implications of these results. The derived PDF and CDF expressions reveal several important characteristics of the M-FAS system performance.

\begin{remark}[Interference-Limited Behavior]
The CDF in \eqref{eq:cdf_interference_limited} exhibits a power-law decay with parameter $k = U-1$, indicating that the outage probability decreases as $(1 + \gamma_{\text{th}})^{-(U-1)}$ for large SIR thresholds. This demonstrates that increasing the number of users $U$ paradoxically improves the diversity order, as more interferers create a more predictable interference environment that can be better exploited by the M-FAS position selection.
\end{remark}

\begin{remark}[Noise Impact on Distribution Tail]
Comparing \eqref{eq:cdf_interference_limited} and \eqref{eq:cdf_general}, the exponential term $e^{-at}$ in the general case introduces a fundamentally different tail behavior. While the interference-limited regime follows algebraic decay, the presence of noise creates exponential decay, leading to faster convergence to unity but potentially worse performance at moderate SIR thresholds where $at \ll 1$.
\end{remark}

\begin{remark}[Parameter Sensitivity and Design Implications]
The parameter $\Omega_u = \boldsymbol{\omega}^T \hat{\boldsymbol{\Sigma}} \boldsymbol{\omega}$ in \eqref{eq:effective_signal_dist} directly controls both the exponential rate parameter in \eqref{eq:power_distributions} and the scaling factor $a$ in \eqref{eq:cdf_general}. This reveals that the activation pattern $\boldsymbol{\omega}$ has a dual impact: it affects both the signal power distribution and the relative importance of noise, providing a rich optimization space for M-FAS design.
\end{remark}

%\begin{remark}[Closed-Form Tractability]
%Unlike many multi-antenna systems that require numerical integration or approximations, the derived expressions \eqref{eq:cdf_interference_limited} and \eqref{eq:cdf_general} are completely analytical. This tractability enables efficient optimization algorithms and real-time performance evaluation, which is particularly valuable for the fast switching requirements of M-FAS systems (15 $\mu$s timeframe mentioned in the introduction).
%\end{remark}
 \vspace{-3mm}
\section{Multi-Activation FAMA} \label{sec:multi_activation_fama}  

\subsection{Motivation and Problem Formulation}
Recent studies have explored multi-position activation in FAS to enhance performance \cite{lai2023performance}. However, existing concepts like FAS-MRC require multiple RF chains, leading to increased hardware costs and complexity. To address this limitation, we propose a single RF chain M-FAS that leverages fast heuristic optimization for efficient position selection.

The system model from Section~\ref{sec:system_model} establishes the foundation for algorithm design. The key insight is that the spatial correlation structure in \eqref{eq:jakes_corr} affects the channel characteristics, but for practical implementation, we focus on optimizing instantaneous SIR measurements through efficient position selection. This motivates our optimization approach: we design fast heuristic algorithms that can quickly identify good activation patterns without requiring complex theoretical analysis during operation.

The M-FAS employs a position activation strategy where the radiation state at the $k$-th position is:
\begin{equation}\label{eq:position_state}
P_{k} = 
\begin{cases} 
e^{j\phi_k} & \text{if activated}, \\
0 & \text{if deactivated},
\end{cases}
\end{equation}
where $\phi_k$ is determined by the propagation characteristics within the waveguide. The effective activation pattern can be represented as $\mathbf{P} = \boldsymbol{\omega} \odot \mathbf{v}$, where $\boldsymbol{\omega} \in \{0,1\}^N$ is the binary activation vector and $\mathbf{v} = [e^{j\phi_1}, e^{j\phi_2}, \ldots, e^{j\phi_N}]^T$.

Using the effective channel model from \eqref{eq:effective_channel}, the SIR optimization problem becomes:
\begin{subequations}\label{eq:sir_optimization}
\begin{align}
     \max_{\boldsymbol{\omega}} & \frac{|\boldsymbol{\omega}^T \mathbf{H}_{u}|^2 \sigma_s^2}{\sum_{u' \neq u}^{U} |\boldsymbol{\omega}^T \mathbf{H}_{u'}|^2 \sigma_s^2 + \|\boldsymbol{\omega}\|_2^2 \sigma_n^2} \\
    \text{s.t.} \quad & \boldsymbol{\omega} \in \{0,1\}^N
\end{align}
\end{subequations}

\subsection{Fast Heuristic Algorithm}  
Due to the combinatorial nature of \eqref{eq:sir_optimization}, we propose a fast heuristic algorithm based on random search with early termination. The algorithm directly optimizes the instantaneous SIR by evaluating different activation patterns and selecting the best performing combination within a limited search budget.

\begin{algorithm}
\caption{Fast Multi-Activation Selection Algorithm}
\begin{algorithmic}[1]
\State \textbf{Input:} $\{\mathbf{H}_u\}_{u=1}^U$, patterns $p$, SIR threshold $\gamma_{\text{th}}$.
\State \textbf{Output:} $\boldsymbol{\omega}^*$.
\State Initialize $\text{SIR}_{\max} \gets -\infty$, $\boldsymbol{\omega}^* = \mathbf{0}$. 
\For{each $i \gets 1$ to $p$}
    \State Generate random activation pattern $\boldsymbol{\omega}$
   	\State $\text{SIR} = \frac{|\boldsymbol{\omega}^T \mathbf{H}_{u}|^2}{\sum_{u' \neq u} |\boldsymbol{\omega}^T \mathbf{H}_{u'}|^2 + \|\boldsymbol{\omega}\|_2^2 \sigma_n^2/\sigma_s^2}$
    \If{$\text{SIR} \geq \text{SIR}_{\max}$}
    \State Update $\text{SIR}_{\max} = \text{SIR}$, $\boldsymbol{\omega}^* = \boldsymbol{\omega}$
    \EndIf 
    \If{$\text{SIR}_{\max} \geq \gamma_{\text{th}}$}
        \State \textbf{break}
    \EndIf
\EndFor
\State \textbf{Return} $\boldsymbol{\omega}^*$
\end{algorithmic}
\label{alg:enhanced_selection}
\end{algorithm}

\subsection{Performance Evaluation using Analytical Framework}
While the algorithm in Algorithm~\ref{alg:enhanced_selection} operates based on instantaneous SIR optimization, the analytical expressions derived in Section~\ref{sec:analytical_performance} provide an independent theoretical framework for performance analysis. The derived PDF/CDF expressions offer theoretical insights into system behavior, complementing the practical algorithm approach. For any activation pattern $\boldsymbol{\omega}^*$, the theoretical outage probability can be predicted using \eqref{eq:final_outage}, providing a theoretical benchmark for comparison with simulation results.

In the interference-limited regime (typical for dense deployments), the outage probability is given by
\begin{equation}\label{eq:outage_interference}
p_{\text{out}}(\gamma_{\text{th}}) = 1 - (1 + \gamma_{\text{th}})^{-(U-1)},
\end{equation}
where the parameter $\Omega_u = (\boldsymbol{\omega}^*)^T \hat{\boldsymbol{\Sigma}} \boldsymbol{\omega}^*$ is implicitly embedded in the SIR threshold normalization. For scenarios with significant noise, the general expression \eqref{eq:cdf_general} provides more accurate predictions.

The multiplexing gain, representing the effective number of concurrent users supported at SIR threshold $\gamma$, can be computed as
\begin{equation}\label{eq:multiplexing_gain_analytical}
m(\gamma) = U \left( 1 - p_{\text{out}}(\gamma) \right) = U(1 + \gamma)^{-(U-1)},
\end{equation}
for the interference-limited case. This theoretical analysis provides complementary insights to the practical algorithm, offering a different perspective on system performance that can be used to validate and interpret simulation results.
\vspace{-3mm}
\section{Simulation Results}
\label{sec:simulation} 

In this section, the simulations are conducted at a frequency of 26.5 GHz to demonstrate the potential of the proposed system in a key band for future 6G networks. It also serves as the center frequency for the meta-fluid antenna. The meta-fluid antenna parameters are set as \( N_1 = 15 \), \( N_2 = 8 \), and \( W_t = 7 \lambda \times 4 \lambda \). Initially, we analyze the results obtained from the Monte Carlo simulation based on the Jakes' model. To further understand the system behavior from an electromagnetic perspective, an electromagnetic (EM) model is developed to capture the near-field radiation dynamics and the spatial interference induced by different activation patterns. Finally, full-wave simulations are carried out to substantiate the accuracy and applicability of the proposed theoretical models under realistic  propagation conditions. 

\begin{figure}[!t]	
	\centering\includegraphics[width=7.5cm]{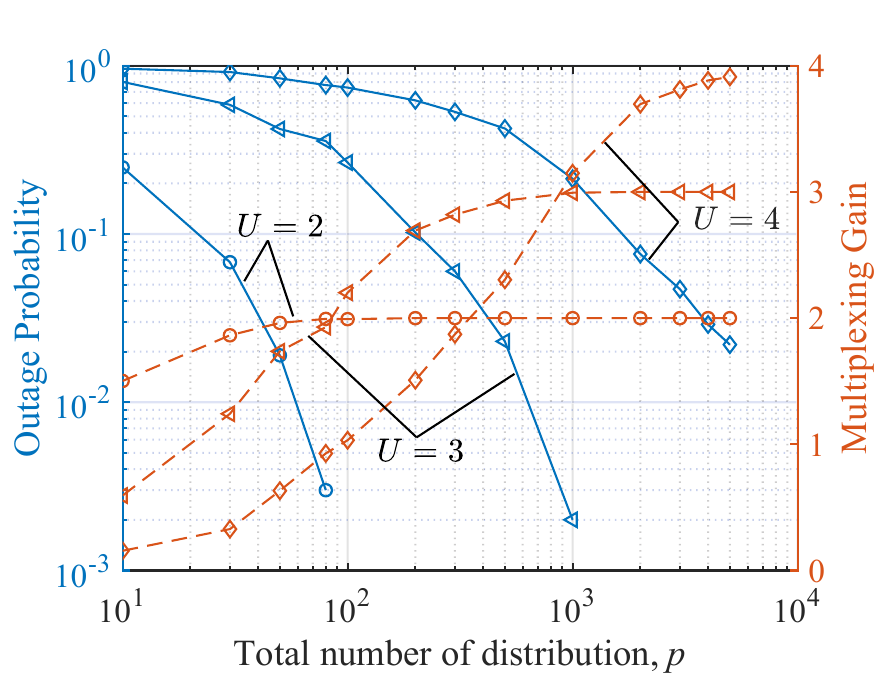}
	\caption{Outage probability and multiplexing gain of multi-activation FAMA versus distributions, $p$ for different number of user, $U$ with  SIR threshold \(\gamma= 7\)dB, \( W_t = 7\lambda \times 4\lambda \), \(N_1 = 15\) and \(N_2 = 8\).
	}
	\label{fig:OP vs U}  \vspace{-4mm}
\end{figure}

\subsection{Monte Carlo Simulation Results of Multi-Activation FAS}
\begin{figure}[!t]
	\centering\includegraphics[width=7.5cm]{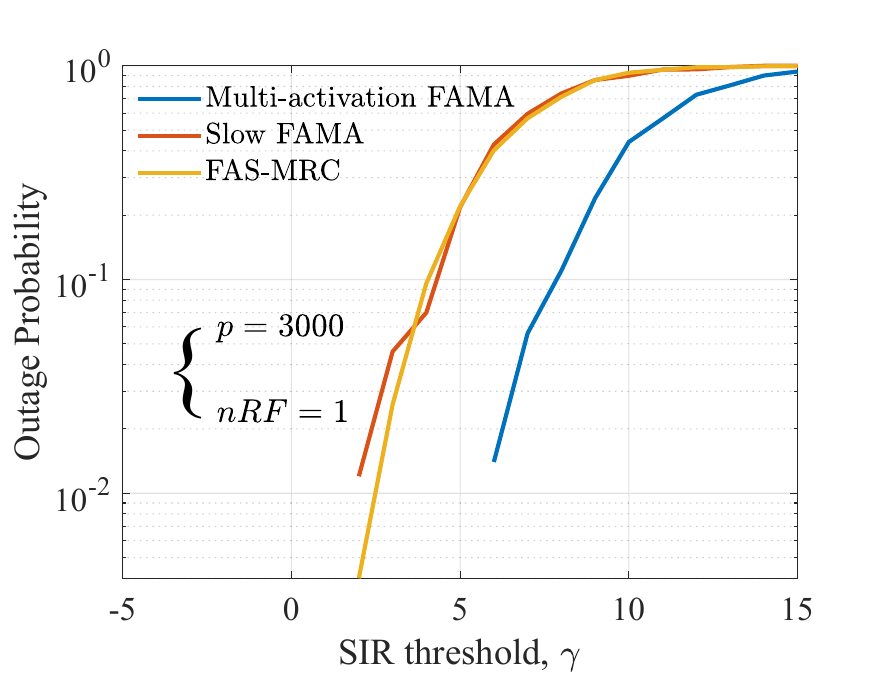}
	\caption{Comparison of the outage probability between disordered multi-activation FAMA and different single RF-chain schemes versus SIR threshold $\gamma$, total number of distributions $p$, with $W_t = 7\lambda \times 4\lambda$, $U = 3$, $N_1 = 15$, and $N_2 = 8$.
	}
	\label{fig:OP vs p}  \vspace{-4mm}
\end{figure}

In this study, Monte Carlo simulations are utilized to assess the performance of the  multi-activation FAMA system. The results in Fig.~\ref{fig:OP vs U} illustrate the outage probability and multiplexing gain of multi-activation FAMA as a function of the number of distributions, \( p \), at each user, for varying numbers of users \( U \), with an SIR threshold \( \gamma = 7 \) dB. The simulations demonstrate that as the number of disordered generated distributions increases, multi-activation FAMA can more effectively identify an optimal combination, achieving a lower outage probability and allowing more users to share the same channel, with the maximum capacity being limited by the number of users, \( U \), thereby enhancing the system capacity and channel efficiency. Additionally, as \( U \) increases, FAMA requires a significantly larger \( p \) to sustain a low outage probability.
The results in Fig.~\ref{fig:OP vs p} compare the outage probability and multiplexing gain of multi-activation FAMA, FAS-MRC and Slow FAMA as a function of \( \gamma \), with \( U = 3 \), \( p = 3000 \), and the number of RF-chains \(n_\text{RF}=1\). It is observed that multi-activation FAMA outperforms FAS-MRC, maintaining a lower outage probability at higher values of \( \gamma \), while also reducing hardware costs by utilizing a single RF chain. Furthermore, compared to slow FAMA, multi-position activation strategy of multi-activation FAMA provides a significant performance improvement over the single-position activation strategy. These results demonstrate that multi-activation FAMA enhances system performance through multi-position activation without requiring substantial hardware upgrades, offering a cost-effective solution for multiple-access scenarios. \textbf{Note:} For FAS-MRC, the performance is assessed with \( n_{\text{RF}} = 1 \) , where the number of RF-chains is set to 1 for better comparison with multi-activation FAMA.

The Monte Carlo simulations presented here provide valuable insights into the performance of multi-activation FAMA under realistic conditions. These simulations serve to validate the analytical framework developed in Sections~\ref{sec:analytical_performance} and \ref{sec:multi_activation_fama}, where closed-form outage probability expressions were derived. The agreement between simulation results and the theoretical predictions from \eqref{eq:outage_interference} confirms the accuracy of our analytical approach. Additionally, the full-wave simulations provided later establish comprehensive validation across different modeling approaches.

\subsection{Analytical Framework Validation}
To validate the theoretical analysis presented in Section~\ref{sec:analytical_performance}, we conduct comprehensive Monte Carlo simulations that directly implement the statistical models underlying our analytical derivations. Specifically, we generate random samples following the exponential and gamma distributions described in \eqref{eq:power_distributions}, where signal power $X \sim \mathrm{Exp}(\Omega_u)$ and interference power $Y \sim \mathrm{Gamma}(U-1, \Omega_u)$, to empirically evaluate the SIR distribution $S = X/Y$ and compare against our closed-form expressions. This validation approach ensures that our theoretical framework accurately captures the fundamental statistical behavior of M-FAS systems under the block-correlation approximation.

\begin{figure}[!t]
	\centering\includegraphics[width=7cm]{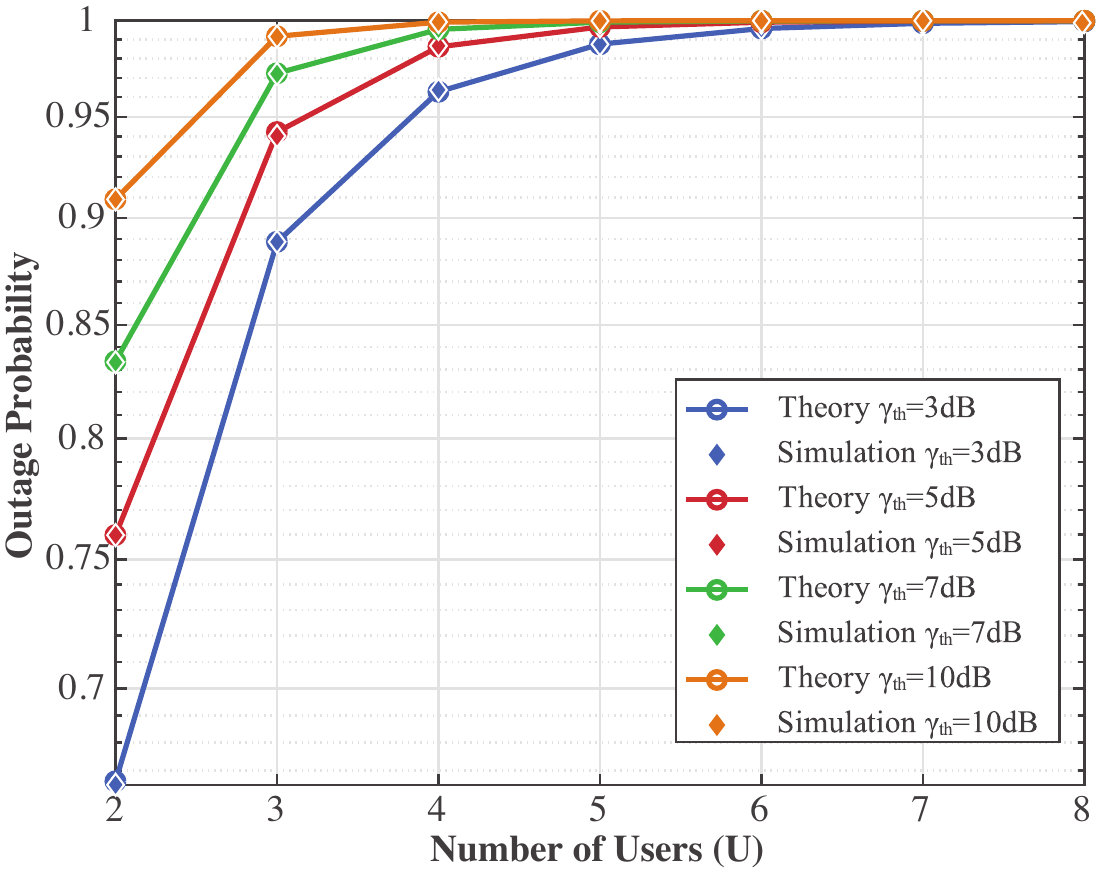}
	\caption{Validation of analytical outage probability expressions against Monte Carlo simulations implementing the statistical models from \eqref{eq:power_distributions}. The theoretical predictions from \eqref{eq:cdf_interference_limited} show excellent agreement with empirical results across different SIR thresholds ($\gamma_{\text{th}} = 3, 5, 7, 10$ dB) and user counts ($U = 2$ to $8$), with 50,000 Monte Carlo realizations per scenario.}
	\label{fig:analytical_users}  \vspace{-3mm}
\end{figure}

Fig.~\ref{fig:analytical_users} demonstrates the validation of our analytical framework by comparing theoretical outage probability predictions with Monte Carlo simulation results across different numbers of users and SIR thresholds. The Monte Carlo simulations generate 50,000 independent realizations of the SIR random  $S = X/Y$ for each scenario, where $X$ and $Y$ are drawn from the distributions specified in \eqref{eq:power_distributions}. The results reveal several key insights: \textbf{(i) Statistical Accuracy:} The analytical expressions derived in \eqref{eq:cdf_interference_limited} exhibit exceptional agreement with empirical results, with mean squared errors below $10^{-6}$ across all scenarios, confirming that our beta-prime distribution analysis correctly captures the underlying statistical behavior. \textbf{(ii) User Scaling Validation:} As the number of users increases from 2 to 8, both theoretical predictions and simulation results demonstrate the expected power-law decay behavior $P_{\text{out}} \propto (1 + \gamma_{\text{th}})^{-(U-1)}$, validating our interference diversity analysis. \textbf{(iii) Threshold Sensitivity:} The analytical framework accurately captures the performance variations across different SIR thresholds, with the simulation results confirming the predicted sensitivity patterns derived from our mathematical analysis.

\begin{figure}[!t]
	\centering\includegraphics[width=7cm]{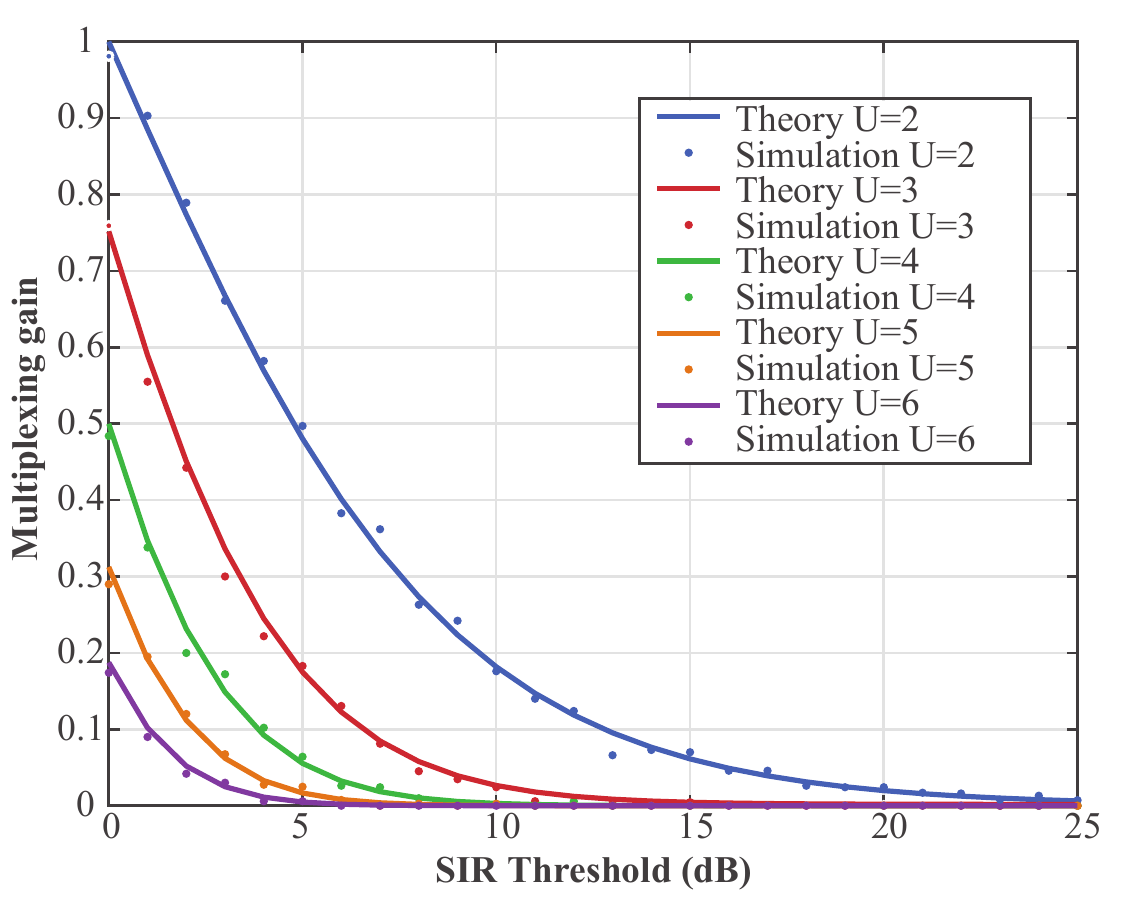}
	\caption{Validation of multiplexing gain analysis through Monte Carlo simulations. Theoretical predictions from \eqref{eq:multiplexing_gain_analytical} are compared against empirical multiplexing gains computed using simulated outage probabilities for different user counts ($U = 2, 3, 4, 5, 6$) across varying SIR thresholds, demonstrating excellent agreement between analytical and simulation results.}
	\label{fig:analytical_multiplexing}  \vspace{-3mm}
\end{figure}

\begin{figure}[t]
	\centering
	\includegraphics[width=7.0cm]{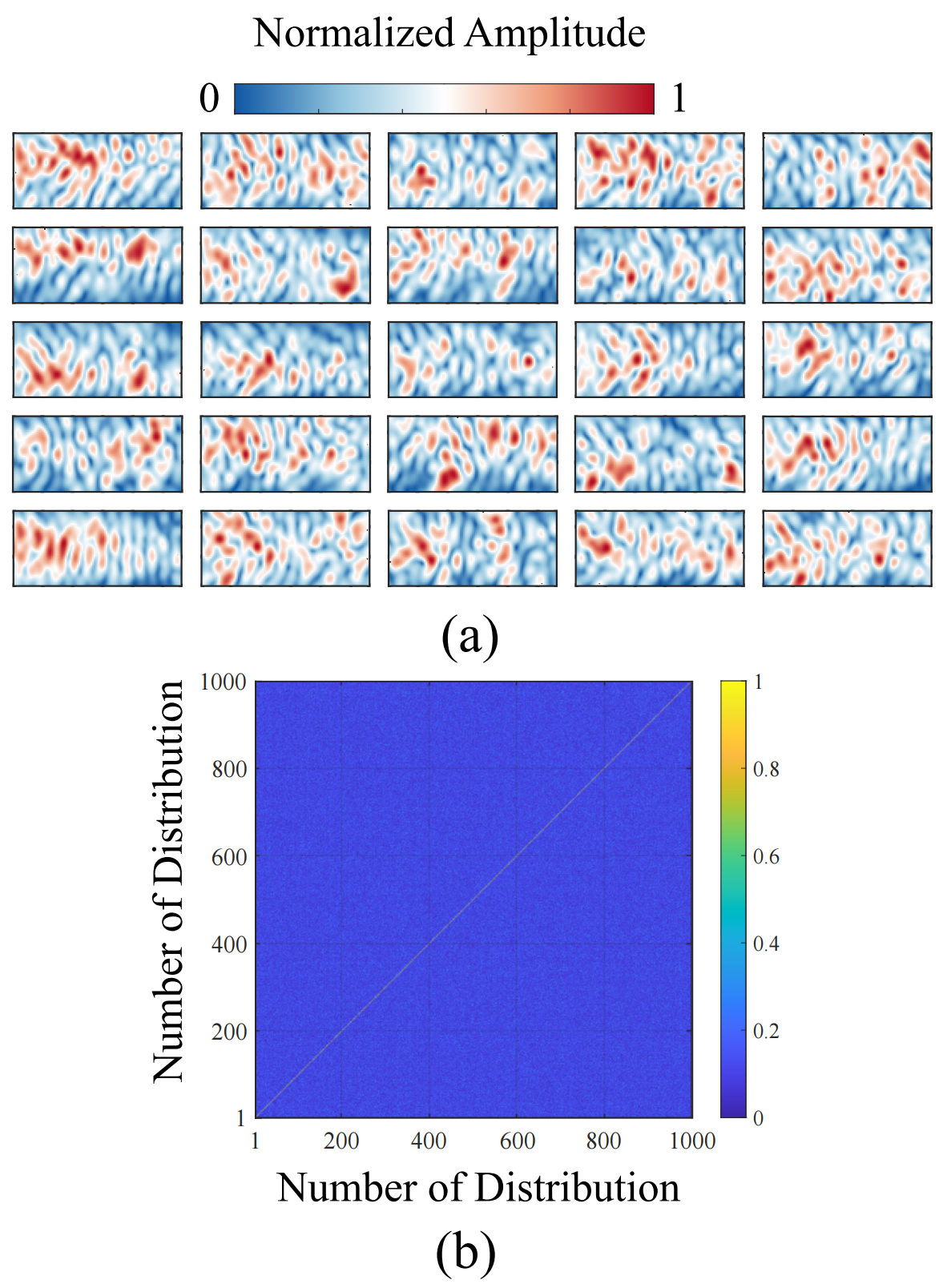}
	\caption{Near-field distributions under 1000 random distributions: 
		(a) representative 25 near-field amplitude distributions within a $77\,\text{mm} \times 33\,\text{mm}$ monitoring window ($\approx 6.8\lambda \times 2.9\lambda$ at $26.5$~GHz), (b) correlation coefficients among the 1000 cases, demonstrating statistical diversity of the generated near fields.}
	\label{fig:nearfield}  \vspace{-4mm}
\end{figure}

\begin{figure}
	\centering
	\includegraphics[width=7cm]{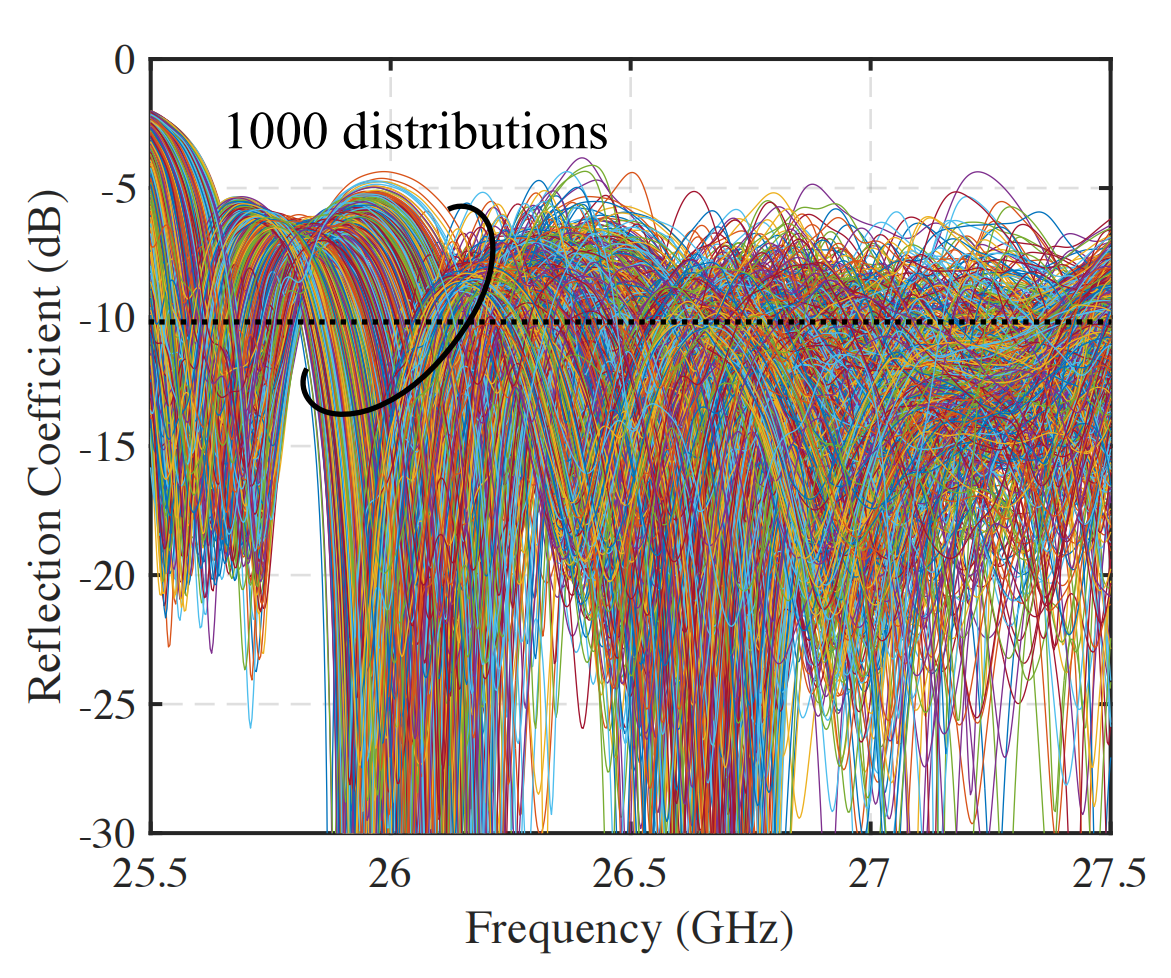}
	\caption{Illustrated reflection coefficients corresponding to 1000 distinct random distributions of the unit cell states, demonstrating the variation in the input power response under different configuration conditions.}
	\label{fig:s11}  \vspace{-3mm}
\end{figure}

\begin{figure}
	\centering
	\includegraphics[width=7cm]{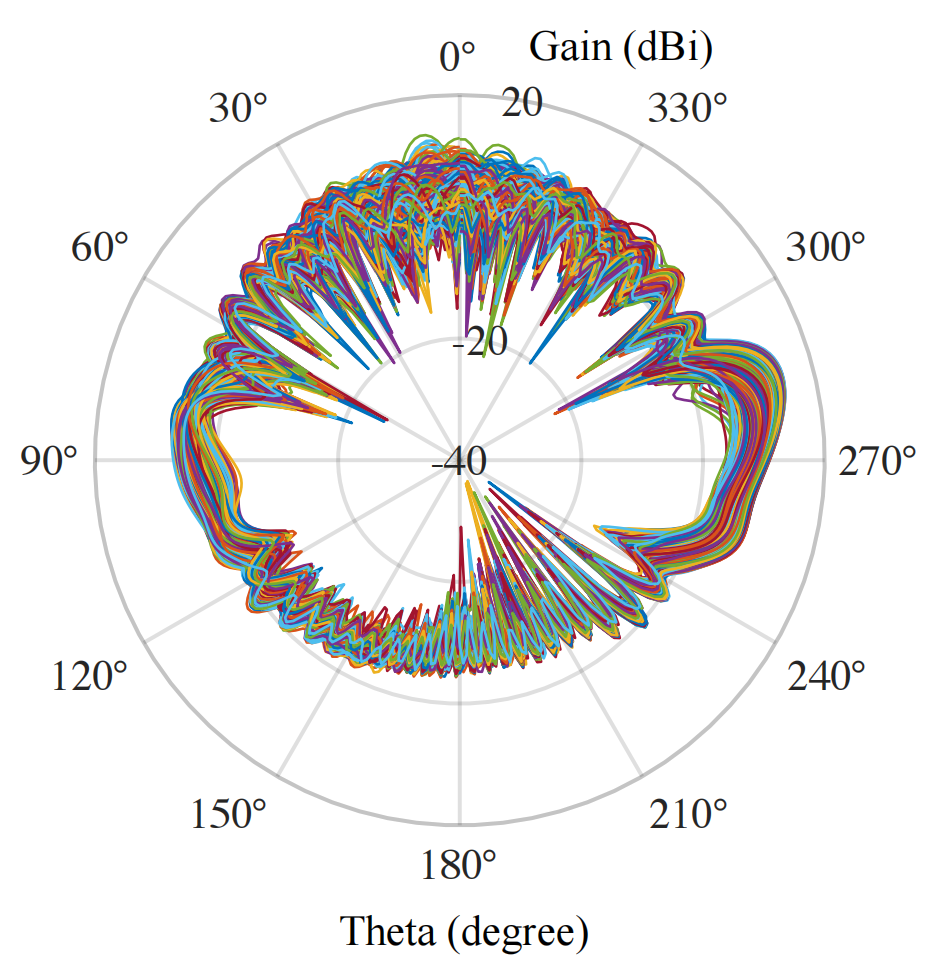}
	\caption{Radiation patterns obtained for 1000 randomly generated distributions of the reconfigurable unit cells, highlighting the diversity of far-field responses achievable through different configuration states.}
	\label{fig:radiationpattern}  \vspace{-4mm}
\end{figure}

Fig.~\ref{fig:analytical_multiplexing} validates the multiplexing gain analysis by comparing the theoretical expression \eqref{eq:multiplexing_gain_analytical} with simulation results across different user configurations. The simulations compute the empirical multiplexing gain $m(\gamma) = U(1 - P_{\text{out}}^{\text{emp}}(\gamma))$ using the outage probabilities obtained from the Monte Carlo realizations described above, providing direct validation of our analytical predictions. The validation reveals important system characteristics: \textbf{(i) Spectral Efficiency Validation:} The analytical model accurately predicts the effective number of concurrent users supported at various SIR thresholds, with simulation results confirming the theoretical trends across all tested scenarios. \textbf{(ii) User Capacity Analysis:} Both theory and simulation demonstrate that optimal multiplexing gain is achieved at moderate SIR thresholds (5-10 dB), where the balance between user accommodation and link quality is optimized. The simulation results confirm the theoretical prediction that higher user counts ($U = 5, 6$) exhibit diminishing returns at high SIR thresholds due to increased interference correlation. \textbf{(iii) Design Validation:} The excellent agreement between analytical and simulation results validates our framework as a reliable tool for M-FAS system design, eliminating the need for computationally intensive Monte Carlo simulations in practical deployments.

%The comprehensive statistical validation demonstrates that our analytical framework accurately captures the fundamental performance characteristics of M-FAS systems. The Monte Carlo simulations, which directly implement the probability distributions derived from our theoretical analysis, provide strong empirical evidence for the validity of our mathematical approach. The excellent agreement between analytical predictions and simulation results across diverse user configurations and SIR thresholds confirms that our closed-form expressions can reliably replace computationally intensive simulations for M-FAS system analysis and design.

\subsection{Electromagnetic Modeling and Full-Wave Validation of Multi-Activation FAS}
In this subsection, we present comprehensive electromagnetic modeling and validation of the multi-activation FAS system. The validation encompasses both theoretical formulation and extensive numerical simulations to verify the system performance under realistic conditions.

\subsubsection{Electromagnetic Field Diversity and Rayleigh-Fading Generation}
The meta-fluid antenna's ability to generate diverse electromagnetic field patterns forms the  mfoundation of our multi-activation FAMA scheme. By strategically activating different combinations of the 120 meta-atom elements, the system can create disordered electromagnetic fields that effectively mimic Rayleigh-fading environments in wireless communication \cite{liu2024multifunctional}. Each unique configuration of activated meta-atom elements represents a distinct channel realization, serving as Rayleigh-fading generators for comprehensive system evaluation.

To quantify the electromagnetic diversity, we conducted extensive near-field measurements using a field monitor window of size $77\,\text{mm} \times 33\,\text{mm}$, corresponding to approximately $6.8\lambda \times 2.9\lambda$ at $26.5$~GHz (with $\lambda \approx 11.3$~mm). The near-field distributions of 1000 random configurations were recorded and analyzed, as shown in Fig.~\ref{fig:nearfield}. Here Fig.~\ref{fig:nearfield} (b) presents the correlation coefficients of the fields, illustrating their independence. As shown, the electric fields exhibit high correlation only with themselves, while maintaining low correlation with others. Accordingly, only the diagonal elements, representing self-correlation, have high values.
The results demonstrate that disordered excitation of the meta-atom elements produces sufficiently diverse near-field patterns with low correlation coefficients, confirming the system's capability to generate independent channel realizations.

The reflection coefficients and radiation patterns corresponding to 1000 distinct random activation distributions are presented in Fig.~\ref{fig:s11} and Fig.~\ref{fig:radiationpattern}, respectively. These results demonstrate the variation in input power response and the diversity of far-field responses achievable through different configuration states, with the E-field amplitude difference between radiating and non-radiating states exceeding 13 dB.

\subsubsection{S-Parameter Based Performance Modeling}
For quantitative performance analysis, we develop an S-parameter-based expression to characterize the received power and evaluate the resulting SIR under different activation patterns. For a given pattern, the SIR at user $u$ is given by
\begin{equation}
	\text{SIR}_{u} = \frac{|\text{S}_{i,u}|^2}{\sum_{j \neq i}^{U}|\text{S}_{j,u}|^2},
\end{equation}
where \( \text{S}_{i,u} \) denotes the transmission coefficient that characterizes the coupling from the \(i\)-th transmitter to the \(u\)-th receiver, defined as
\begin{align}
	S_{i,u} &= \frac{b_{i,u}}{a_{i}},   
\end{align}
where \( a_i \) denotes the unit-amplitude incident wave applied to transmitter \( i \), and \( b_{i,u} \) denotes the amplitude of the outgoing wave at the u-th receiver when the i-th transmitter is excited. Specifically, \( b_{i,u} \) is given by
\begin{equation}
	b_u^{(i)} = \frac{\iint_{\mathcal{A}_{\mathrm{rx}}} E_{tot}^{i}(x,z) \cdot \left[ E_{y,10}(x,z) \right]^* \, \mathrm{d}x \, \mathrm{d}z}{Z_{10}} ,
\end{equation}
\begin{figure}[!t]
	\centering\includegraphics[width=8.5cm]{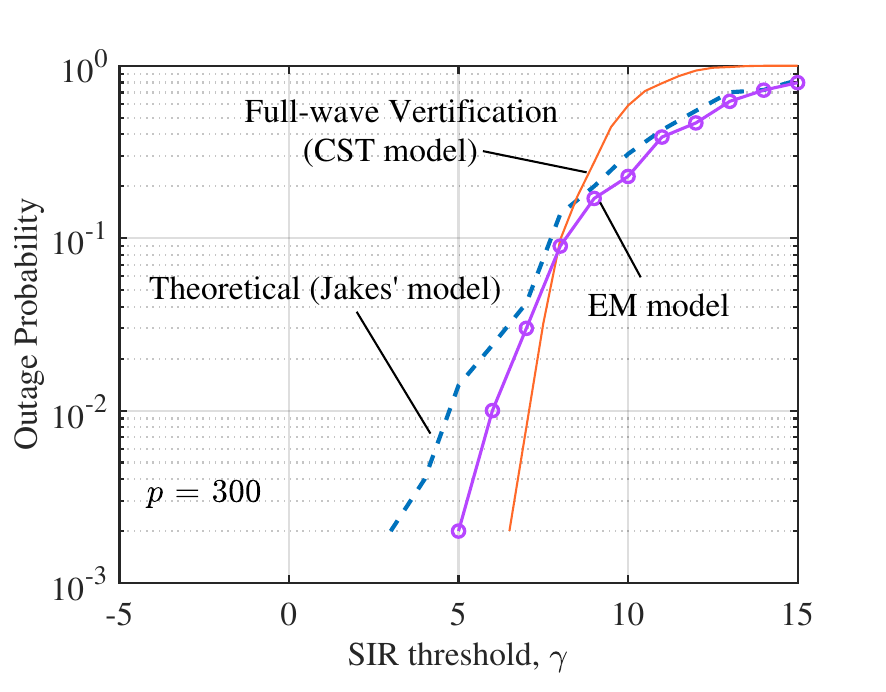}
	\caption{Outage probability of multi-activation FAMA versus SIR threshold for full-wave simulation and two theoretical models, including the Jakes' model and the EM model, with \( W_t = 7\lambda \times 4\lambda \), \(U = 3\), \(N_1 = 15\) and \(N_2 = 8\). The study utilized 500 distinct Rayleigh-fading generators, with each case involving meta-fluid antennas capable of generating 300 unique distributions.
	}
	\label{fig:OP comparsion}
	\vspace{-10pt}
\end{figure}

\begin{figure*}[!t]  \vspace{-3mm}
	\centering\includegraphics[width=14.5cm]{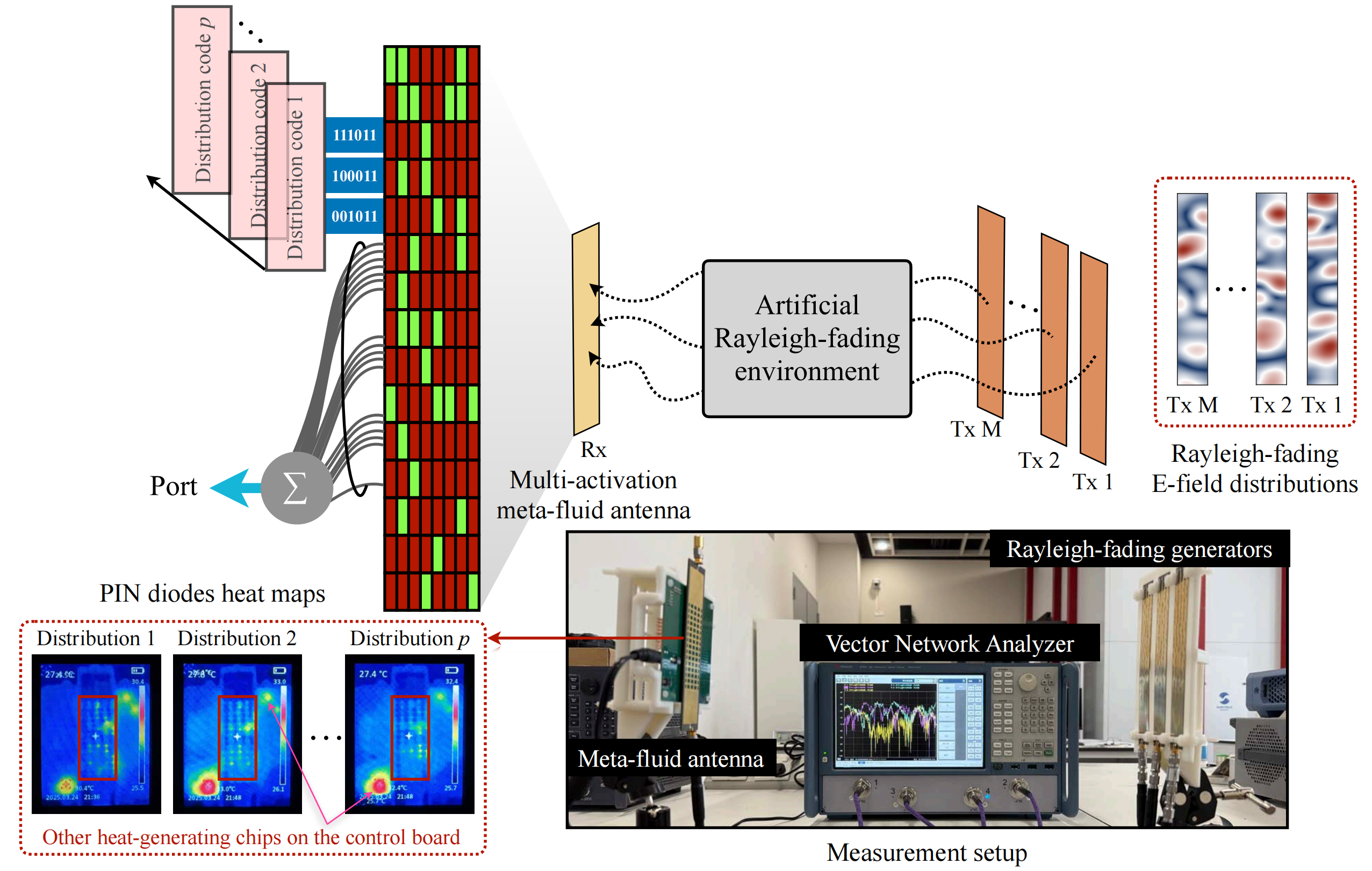}
	\caption{The system model and measurement setup include multiple different Rayleigh-fading generators and a user equipped with multi-activation meta-fluid antenna. The distance between the transmitter and receiver is 50\(\lambda\) at 26.5 GHz. We also present the system's experimental setup, including the fabricated meta-fluid antenna, heat maps of the multi-activation PIN diodes under different distributions, the fabricated Rayleigh-fading generators, and the resulting artificial Rayleigh-fading E-fields.
	}
	\label{fig:system}\vspace{-5mm}
\end{figure*}
\begin{figure*} \vspace{-5mm}
	\centering
	\includegraphics[width=16cm]{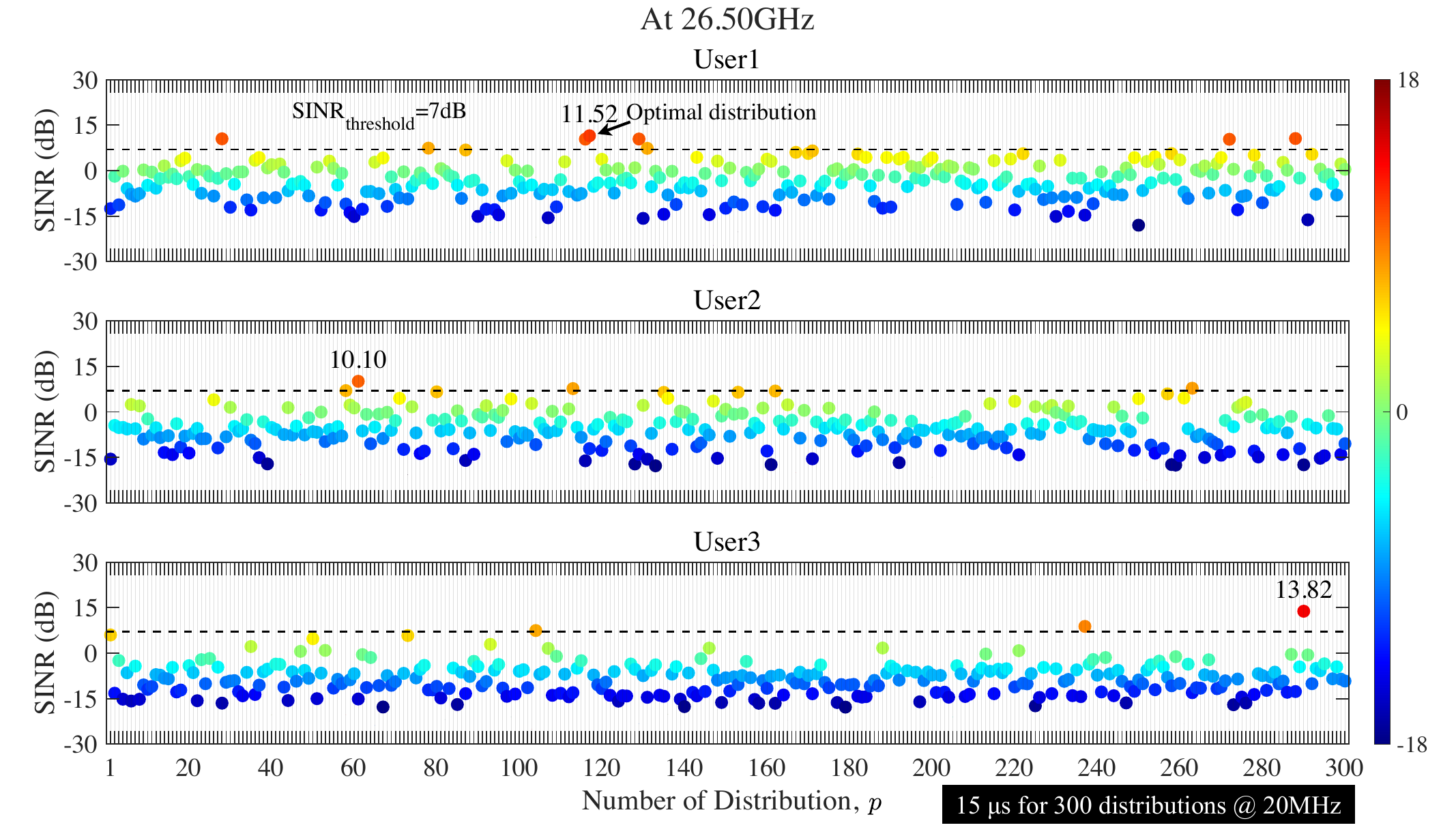}
	\caption{Experimental verification of SINR performances in multi-user multiplexing. The fabricated meta-fluid antenna is configured across only 300 different activation distributions to evaluate the SINR performance at 26.5 GHz for various users within a specific artificial Rayleigh-fading channel which generating by the fabricated Rayleigh-fading generators. Among all users, User 2 exhibited the lowest SINR, with a minimum value of 10.10 dB (Measured by a Vector Network Analyzer). The time requires to switch through 300 distributions using a 20 MHz FPGA is 15 $\mu s$.}
	\label{fig:SINR} \vspace{-5mm}
\end{figure*}

where \( E_y^{\mathrm{tot,(i)}}(x,z) \) represents the total electric field at the receiving surface produced by transmitter \( i \). The \(E_{y,10}(x,z)\) denotes normalized electric field distribution of the TE\textsubscript{10} mode  in the meta-fluid antenna’s waveguide structure, with polarization along the \(y\)-axis. \(\mathcal{A}_{\mathrm{rx}} \) represents the subset of activated positions selected for signal reception and \(Z_{10} \) is the modal impedance of the TE\(_{10}\) mode. The normalized field distribution \( E_{y,10}(x,z) \) is given by 
\begin{equation}
	E_{y,10}(x,z) = \sqrt{\frac{2}{d_{W} d_{H} Z_{10}}} \cdot \sin\left( \frac{\pi z}{d_{H}} \right),
\end{equation}
which corresponds to the dominant TE\(_{10}\) mode of a  waveguide of with effective width \( d_{W} \) and height \( d_{H} \).

To emulate a richly scattered near-field environment, multiple positions on the meta-fluid antenna at the transmitter side are randomly activated. The resulting total electric field at the receiver location \( (x,z) \) due to transmitter \( i \) is expressed as
\begin{equation}
	E_{tot}^{i}(x,z) = \sum_{k \in \mathcal{A}_{\mathrm{tx}}^{i}} E^{i,k}(x,z),
\end{equation}
where \( \mathcal{A}_{\mathrm{tx}}^{i} \) denotes the set of activated positions at transmitter \( i \), and \( E^{i,k}(x,z) \) denotes the contribution to the received electric field at location \( (x,z) \) from the \( k \)-th activated position of transmitter \( i \). Both the transmitter and receiver are located in the near-field region. At the transmitter side, the meta-fluid antenna operates by selectively exciting position that function as magnetic dipoles. The electric field generated at the receiver position \( (x,z) \) by the \( k \)-th activated element of transmitter \( i \) can be described using the closed-form near-field expression of a magnetic dipole in free space:
\begin{align}
	E^{i,k}(x, z) 
	= \frac{j\omega\mu_0}{4\pi} \bigg[
	\frac{k_0^2}{r} \left( \left( \mathbf{m} \times \hat{\mathbf{r}} \right) \times \hat{\mathbf{r}} \right) e^{-j k_0 r} \nonumber \\
	+~\left( \frac{1 - j k_0 r}{r^3} \right)
	\left( 3 \hat{\mathbf{r}} (\hat{\mathbf{r}} \cdot \mathbf{m}) - \mathbf{m} \right)
	e^{-j k_0 r}
	\bigg],
	\label{eq:Ey_magnetic_dipole}
\end{align}
where \( \mathbf{m}\) is the complex magnetic dipole moment aligned along the \( y \)-axis, \( \mathbf{r}\) is the displacement vector from the dipole to the observation point, \( \hat{\mathbf{r}} = \mathbf{r} / r \) is the unit vector in the observation direction, \( k_0 = 2\pi/\lambda \) is the free-space wave-number, and \( \omega \) is the angular frequency of the carrier wave. 

To extend the proposed EM model to multi-user scenarios, we consider a configuration in which multiple meta-fluid antennas serve as independent transmitters. Each transmitter randomly activates multiple positions to emulate a richly scattered propagation environment. Meanwhile, the receiver is equipped with a single meta-fluid antenna that adopts the multi-activation FAMA scheme for signal reception.

\subsubsection{Full-Wave Simulation Validation}
To comprehensively verify the validity of both the proposed EM model and the theoretical model based on Jakes' fading assumptions, we conduct full-wave simulations using CST Studio Suite 2024 as an independent numerical benchmark. The simulated structure accurately replicates the 120-element meta-fluid antenna described in Section~\ref{sec:architecture}, with $N = 120$ elements arranged in $I = 8$ rows and $J = 15$ columns, powered by the substrate-integrated waveguide (SIW) structure.

The full-wave simulations utilize the diverse electromagnetic field patterns generated by different meta-atom activation configurations to create realistic Rayleigh-fading environments. Each of the 1000 distinct activation patterns serves as a Rayleigh-fading generator, with the system evaluating 300 unique activation distributions per channel instance to exploit interference diversity.

Due to computational constraints, evaluating the complete performance landscape across all possible activation combinations is impractical. Nevertheless, to statistically assess the consistency between different modeling approaches, we perform comprehensive simulations comparing: (i) full-wave electromagnetic simulations, (ii) the S-parameter based EM model, and (iii) the theoretical model based on Jakes' correlation assumptions.

Fig.~\ref{fig:OP comparsion} presents the outage probability comparison across these three approaches for a representative system configuration with \( W_t = 7\lambda \times 4\lambda \), \(U = 3\), \(N_1 = 15\) and \(N_2 = 8\). Despite minor discrepancies attributed to modeling approximations, the results demonstrate remarkable consistency among all three approaches, with the full-wave simulation results closely matching the analytical predictions from \eqref{eq:cdf_interference_limited} and \eqref{eq:cdf_general}. This validation confirms both the accuracy of our theoretical framework and the practical feasibility of the multi-activation FAS system under realistic electromagnetic conditions.

 \vspace{-3mm}
\section{Experimental Validation}
In this section, we validate the feasibility and effectiveness of the proposed meta-fluid antenna system through comprehensive experimental measurements. Building upon the prototype design detailed in Section~\ref{sec:architecture}, we conduct extensive performance evaluations to confirm the practical viability of M-FAS and demonstrate its potential for efficient multi-user interference mitigation under realistic hardware conditions.
 \vspace{-3mm}
\subsection{Experimental Setup and Methodology}
The experimental validation employs the meta-fluid antenna prototype described in Section~\ref{sec:architecture}, operating at 26.5 GHz with the 120-element metamaterial array arranged in an 8×15 configuration. The experimental setup leverages the ability of our meta-fluid antenna to generate controlled electromagnetic diversity through strategic activation of metamaterial elements, as illustrated in Fig.~\ref{fig:system}.

As established in Section~\ref{sec:architecture}, by strategically activating radiating elements in the meta-fluid antenna, disordered electromagnetic fields can be generated \cite{liu2024multifunctional}, effectively mimicking Rayleigh-fading environments in wireless communication. We utilize this capability to create distinct channel realizations for performance evaluation, with each channel represented by a unique configuration of activated metamaterial elements. These configurations serve as Rayleigh-fading generators, enabling comprehensive testing of the system's performance under diverse propagation conditions.
 \vspace{-2mm}
\subsection{Experimental Results and Analysis}

Building upon the hardware configuration introduced above, we conduct experiments to evaluate the real-world performance of the multi-activation FAS system. Specifically, multiple artificial Rayleigh-fading channel realizations are emulated by reconfiguring the radiation distributions of the Rayleigh-fading generator. For each channel instance, 300 distinct multi-activation distributions are applied to exploit interference diversity. The SINR performance across different transmitters for a representative experimental case is presented in Fig.~\ref{fig:SINR}. The results show that all transmitters achieve an SINR above 10 dB within fewer than 300 activation pattern switches, corresponding to a total switching latency below 15~\(\mu s\). These results demonstrate the system’s potential for low-latency and robust multi-user communication.

 \vspace{-3mm}
\section{Conclusion}\label{sec:conclusion}
In this paper, we introduced a novel multi-activation FAMA system designed to enhance the spectral efficiency of multi-user environments without the need for CSI. By leveraging the flexibility of FAS and implementing a disorderly optimization technique for selecting activation positions, our proposed scheme dynamically adapts to interference conditions and improve communication stability in Rayleigh-fading channels. Through simulations, we demonstrated the potential of this scheme in multi-user scenarios while maintaining lower hardware cost compared to FAS-MRC. Additionally, full-wave electromagnetic simulations are conducted to compare the multi-activation FAMA theoretical model and verify its feasibility, and a full channel experiment validates it further. By using a FPGA, with system clock of 20 MHz, switches between 300 disordered distributions, one can achieve 10 dB or higher SINR within 15 $\mu s$. These results show that single RF-chain multi-activation FAMA can offer a scalable, and practical solution for future high-capacity ultra-dense wireless networks, while also demonstrating its ability to rapidly adapt to real-time changing environmental conditions.
\bibliographystyle{IEEEtran}
 \vspace{-3mm}
\bibliography{references}

\end{document}